\documentclass[twocolumn,prl,aps]{revtex4}

\usepackage[dvips]{graphicx}

\begin{document}

\title{Hybrid Goldstone modes in multiferroics}

\author{
S. Pailh\`es$^1$, X. Fabr\`eges$^1$, 
L. P. R\'egnault$^2$, 
L. Pinsard-Godart$^3$, 
I. Mirebeau$^1$, 
F. Moussa$^1$,
M. Hennion$^1$, and 
S. Petit$^1$}

\affiliation{
$^1$ Institut Rayonnement Mati\`ere de Saclay, 
Laboratoire L\'eon Brillouin, CEA-CNRS UMR 12, CE-Saclay, 
F-91191 Gif-sur-Yvette, France \\
$^2$ Institut Nanosciences et cryog\'enie, 
CEA-Grenoble, DRFMC-SPSMS-MDN, 
17 rue des Martyrs, F-38054 Grenoble Cedex 9, France \\
$^3$ Institut de Chimie Mol\'eculaire et des Mat\'eriaux 
d'Orsay, Laboratoire de Physico-Chimie de l'Etat Solide, 
UMR CNRS 8182. B\^at 410, Universit\'e Paris Sud. 
F-91405 Orsay Cedex, France
}

\date{\today}

\begin{abstract}

By using polarized inelastic neutron scattering measurements, 
we show that the spin-lattice quantum entanglement in mutliferroics 
results in hybrid elementary excitations, involving spin and 
lattice degrees of freedom. These excitations can be considered as 
multiferroic Godstone modes. We argue that the Dzyaloshinskii-Moriya 
interaction could be at the origin of this hybridization. 
\end{abstract}


\maketitle

Spontaneous symmetry breaking is a powerful concept at 
the basis of many developments in physics, especially 
in condensed matter and high energy physics. The low 
symmetry phase is described by an order parameter 
associated with low energy and long wavelength excitations, 
restoring the original high temperature phase symmetry. 
These Goldstone modes \cite{1} are nothing but longitudinal 
and transverse phonons in solids, or spin waves in magnets. 
In case of multiferroic materials, two order parameters, 
namely the ferroelectric polarization and the magnetization, 
coexist and are strongly coupled by a spin-lattice 
interaction \cite{2,3}. As a result of this entanglement, 
the multiferroic Goldstone modes are expected to be new spin 
and lattice hybrid excitations called electromagnons 
\cite{4,5,6,7}. While their 
existence has been theoretically predicted for a 
long time \cite{8,9}, their dual nature as both spin 
and lattice excitations makes them challenging to 
observe and study experimentally. Due to their dipole 
electric activity, they first could be detected by optical 
measurements \cite{4,8,9} : evidence for their existence 
has been recently reported in different orthorhombic multiferroics, 
namely $GdMnO_3$, $TbMnO_3$ \cite{10}, 
$Eu_{0.75}Y_{0.25}MnO_3$ \cite{11}, $DyMnO_3$ \cite{12}, 
$YMn_2O_5$ and $TbMn_2O_5$ \cite{13}. However, their 
magnetic counterpart is still not clearly evidenced.
Morevover, as optical techniques probe the zone center, the 
shape of their dispersion and thus the underlying 
mechanism responsible for the hybridization, is still unknown. 
In this letter, we report polarized inelastic neutron 
scattering experiments performed on the particular case of 
hegagonal $YMnO_3$ to unravel the spin-lattice dual nature 
of these hybrid modes. 
Moreover, as neutron scattering allow a global survey of 
the reciprocal space, we report the dispersion of these 
hybrid modes throughout the Brillouin zone. This result is 
discused in the framework of the dynamical magnetoelectric 
coupling theory, where the Dzyaloshinskii-Moriya interaction 
plays a central role. 

$YMnO_3$ becomes ferroelectric below 900K, with an 
electric polarization along the c-axis, due to 
alternatively long and short yttrium-oxygen bonds 
(parallel to the c-axis) \cite{14,15}. Despite a 
strong geometric frustration, the Mn spin order 
below the N\'eel temperature $T_N=75K$ in a classical 
triangular arrangement \cite{16,17,18} (Curie-Weiss 
temperature $\theta ~ 500K$). The magnetic and 
electric order parameters strongly interact, as 
recently shown in reference \cite{19}, claiming 
the existence of a giant magneto-elastic coupling".

\begin{figure}[t]
\centerline{
\includegraphics[width=8 cm]{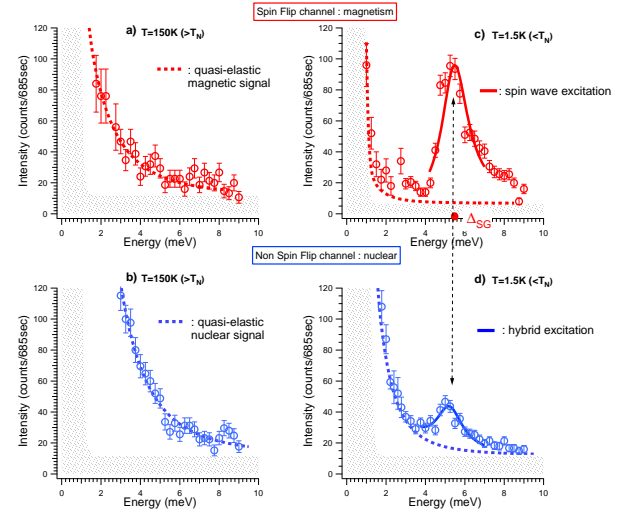}
}
\caption{(Color online) 
Dynamical magnetic and nuclear changes from the 
paramagnetic phase to the Néel ordered state. 
Figures (a) to (d) report energy scans in the SF 
(red labelled) and NSF (blue labelled) channels 
at the wave vector Q=(0,0,6) measured in the 
paramagnetic phase (left column) and in the N\'eel 
state (right column). Red and blue lines are fits 
to the data including the spin waves and the nuclear 
mode for low temperature scans.
}
\label{fig1}
\end{figure}

Through the comprehensive investigation of hybrid 
spin and lattice excitations, inelastic neutron 
scattering on a triple axis spectrometer combined 
with longitudinal polarization analysis (LPA) 
\cite{20,21} offer an efficient way to determine 
hybrid modes. Indeed, this technique allows to 
measure separately the spin-spin and nuclear-nuclear 
correlation functions in different channels (called 
spin flip SF and non spin flip NSF respectively. 
It is worth noting that the magnetic cross section 
probes spin fluctuations perpendicular to the wave 
vector Q only. On the contrary, the nuclear cross 
section probes the atomic displacements parallel 
to Q. Measurements were carried out on the IN22 triple 
axis spectrometer at the Institute Laue Langevin (ILL, G
renoble, France). The sample was aligned in the 
scattering plane (100),(001) such that momentum t
ransfer of the form Q=(H,0,L) in units of reciprocal 
lattice wave vectors were accessible and mounted into a
n ILL-type orange cryostat (1.5K-300K). All data  
were obtained with a fixed final wave vector of 
$2.662 \AA^{-1}$ providing an energy resolution less 
than 1meV. Heussler crystals were used as analyzer 
and monochromator, together with a flipper of Mezei 
to reverse the spin of the scattered neutrons. The 
elastic and inelastic measurement of the polarisation 
efficiency (flipping ratio), as determined from different 
magnetic Bragg peaks (100),(105),(003) and from the 
magnon signal at Q=(-0.4 0 6) and (1 0 1), was of 
about 16 (elastic) and 14 (inelastic signal). The 
amplitude of the expected leak from the SF channel 
to the NSF channel is therefore less than 7\% of the 
SF intensity. Using an unpolarized beam (PG monochromator), 
we measured a polarization parallel and perpendicular 
in- and out-of-plane 
to the wave vector less than $10^{-2}$ excluding the 
presence of chiral terms and of the nuclear-magnetic 
interference terms \cite{20,21}. 

\begin{figure}[t]
\centerline{
\includegraphics[width=8 cm]{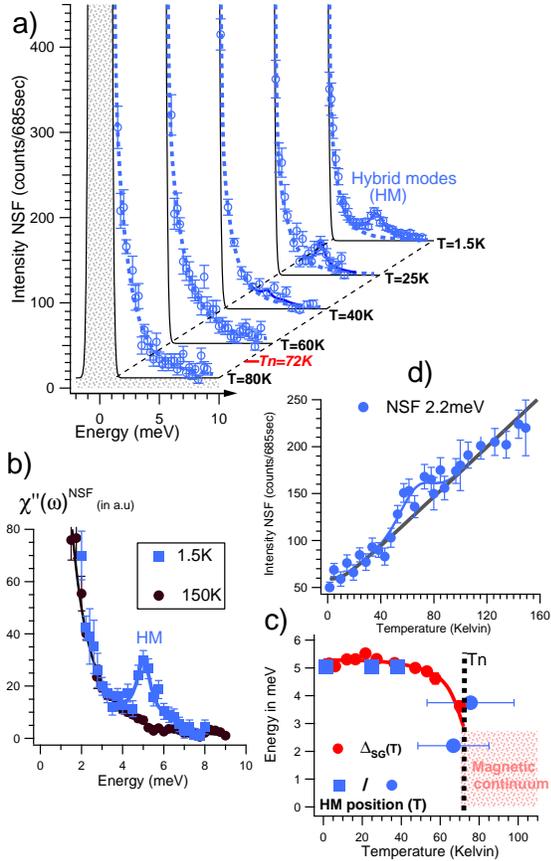}
}
\caption{(Color online) 
Nuclear mode at the zone center in the Néel phase. 
(a) Energy scans in the NSF channels at the wave 
vector Q=(0,0,6) for different temperatures above 
and below $T_N$. Below $T_N$, the additional nuclear 
excitation starts to form from the 40K (blue line). 
(b) Background subtracted and detailed balance factor 
divided neutron intensity of a). (c) Temperature 
dependences of the spin gap (red points, measured 
with unpolarized neutrons) and of the nuclear mode 
(blue points) positions. (d) Temperature dependence 
of the NSF inelastic intensity at 2.75 meV. The gray 
line corresponds to the detailed balance factor. 
}
\label{fig2}
\end{figure}

Energy scans at the wave vector Q=(0,0,6) in both 
channels measured above and below $T_N$ are depicted 
on the panel B of the Figure \ref{fig1}. On the one 
hand, the strong magnetic quasi-elastic signal observed 
at 150K in the SF channel confirms the presence of 
strong spin-spin correlations in the paramagnetic 
phase arising from the geometrically frustrated Mn 
moments \cite{16,17}. Below $T_N$, 3 almost doubly 
degenerate spin wave modes are known to rise up 
\cite{16,22,23}. At the zone centre, they are 
characterized by a spin gap $\Delta_{SG}$, typical 
of a magnetic anisotropy. Upon cooling below $T_N$, 
the magnetic long range order develops and $\Delta_{SG}$ 
gradually shifts to higher values (Figure 2c), reaching 
its maximum of about 5.3 meV around 40K. On the other 
hand, the NSF intensity shows at 150 K a quasi-elastic 
signal as well. At 1.5K, the NSF data demonstrate the 
emergence of an additional inelastic nuclear mode. The 
energy of this mode coincides with the spin gap $\Delta_{SG}$, 
pointing out its close connection with the spin subsystem. 
To understand the temperature dependence of both 
quasi-elastic and inelastic modes, NSF energy scans 
have been performed at different temperatures around 
$T_N$ (Figure \ref{fig2}a). From these NSF data, we 
extract the lattice susceptibilities, presented on 
Figure \ref{fig2}b. This analysis shows that the additional 
mode rises upon cooling on the top of the quasi-elastic 
signal as a supplementary intensity. Below 40K, the energy 
of this nuclear mode is found to follow the spin gap 
$\Delta_{SG}$ (Figure \ref{fig2}c). Above 40K, its 
presence becomes however hard to distinguish. To 
overcome this difficulty, we measured the NSF intensity 
at Q=(0,0,6) for sampling temperatures at 2.2 meV, well 
below the low temperature energy of the nuclear mode 
(Figure \ref{fig2}d). According to the above analysis, 
the quasi-elastic signal is expected to give a contribution 
proportional to the standard detailed balance factor (grey
 line). Figure \ref{fig2} shows however subtle deviations 
from it in the intermediate region surrounding $T_N$ (blue 
line), that we attribute to the raising of the nuclear mode 
starting from the lowest energies. These measurements allow 
estimating the energy of the nuclear mode for temperatures 
higher than 40K. Finally, we conclude that both spin gap a
nd nuclear mode energies follow the same temperature 
dependence from 1.5 K till $T_N$. Since the nuclear cross 
section probes fluctuations along Q=(0,0,6), which is 
parallel to the c-axis, we attribute the nuclear quasi-elastic 
signal to relaxational vibrations along this particular direction. 
The nuclear mode corresponds to collective vibrations along 
the same c-axis. Moreover, since it is found at a zone center, 
it corresponds to vibrations within the decoration of the unit 
cell. Now the question is: how do such internal motions 
propagate through the crystal? 

\begin{figure}[t]
\centerline{
\includegraphics[width=8 cm]{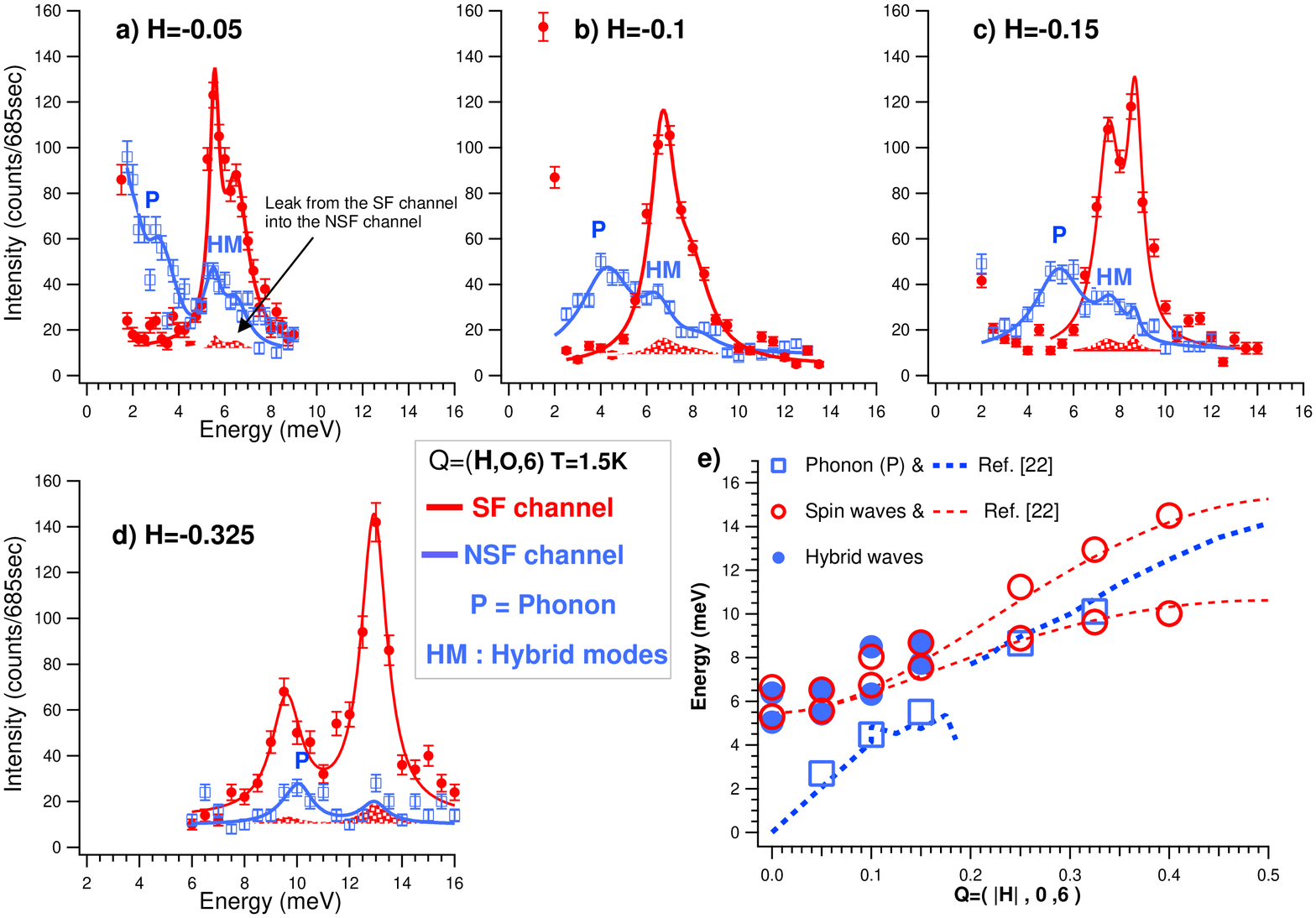}
}
\caption{(Color online) 
Nuclear and magnetic energy dispersions. Energy scans 
taken at 1.5K for different wave vectors Q=(H,0,6) in 
both SF (full points and red fits) and NSF (open points 
and blue fits) channels. The H component is indicated 
in the top left corner of each panel. The dotted line 
indicates the position of the nuclear mode (see text). 
The lowest right panel depicts the dispersions of the 
different low energy excitations : measured (filled 
red points) and calculated (dotted red lines) spin 
waves,  transverse acoustic phonon6 (blue square) 
and hybrid modes (open blue circles). 
}
\label{fig3}
\end{figure}

To address this question, it is essential to determine the 
dispersion of the additional nuclear mode, by repeating the 
same measurements for different Q=(H,0,6) with varying H 
(Figure \ref{fig3}). We first determine the dispersion of 
the modes which are trivially expected, namely the spin 
wave modes (red points in SF channel on Figure \ref{fig3}) 
and the transverse acoustic phonon22 (denoted by "P on 
Figure \ref{fig3}). Amazingly, the NSF data demonstrate 
unambiguously the existence of additional nuclear modes 
all along the low energy spin wave dispersions from 
Q=(0,0,6) to Q=(0.25,0,6) (denoted by "HM on Figure 
\ref{fig3}). Their intensity is found to decrease as the 
wave vector increases. For instance, at Q=(0.325,0,6), the 
nuclear intensity measured in the NSF channel becomes 
comparable to the expected leak of the intensity from the 
SF channel. Figure \ref{fig3}e evidences the fact that the 
dispersions of these nuclear modes match the spin wave 
dispersions for a wide range of wave vectors. It therefore 
appears obvious that they must be attributed to collective 
vibrations within the decoration that propagate through the 
crystal in the N\'eel state by hybridizing with the spin waves. 

\begin{figure}[t]
\centerline{
\includegraphics[width=8 cm]{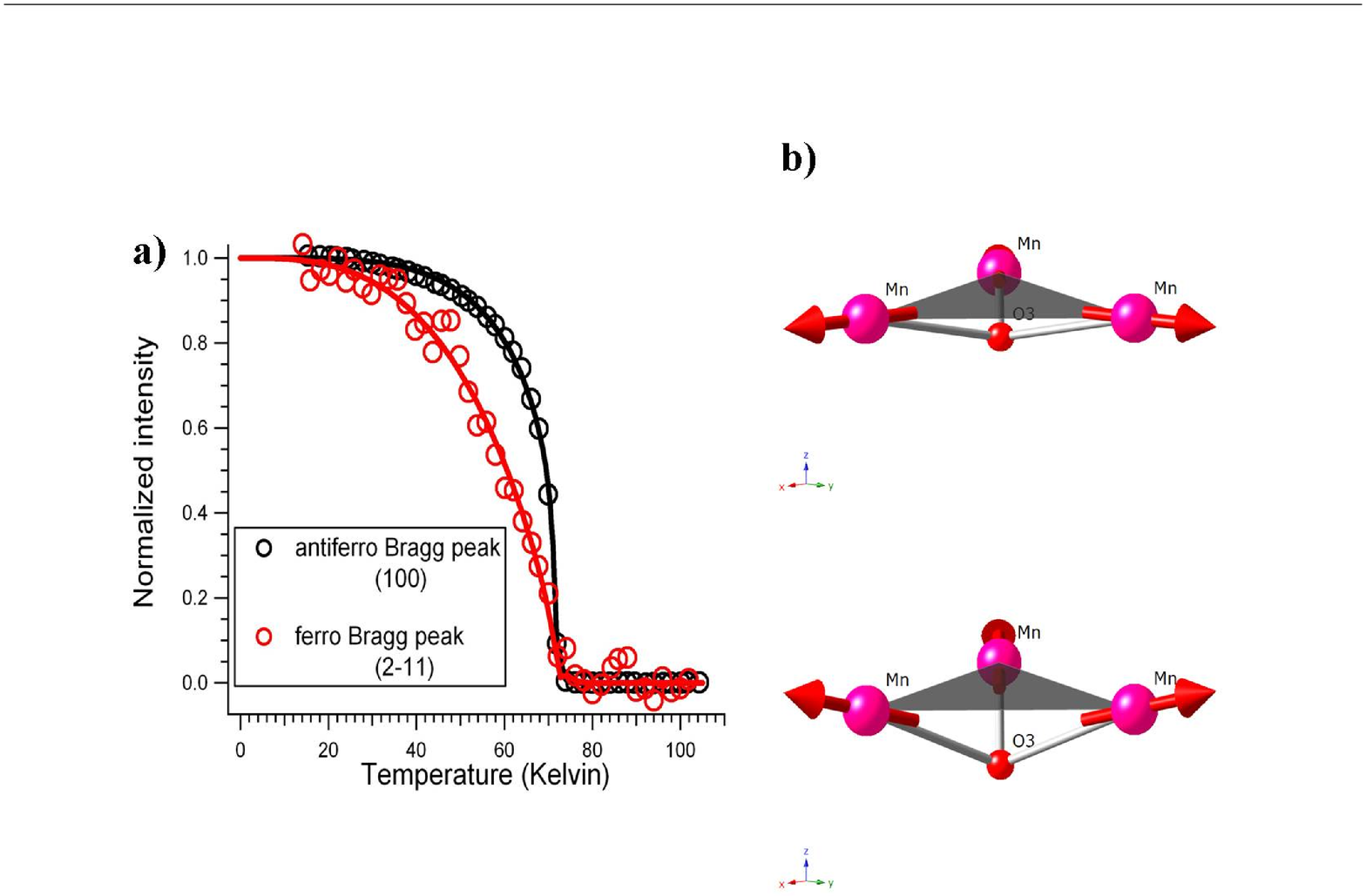}
}
\caption{(Color online) 
Umbrella scenario. Left figure (a) displays the 
temperature dependence of the (100) antiferomagnetic 
bragg peak and of the (2-11) Bragg peak (unpolarized 
neutron).The "umbrella mechanism discussed in the 
text is sketched on the right panel (b). It evidences 
the existence of a small ferromagnetic component 
perpendicular to the Manganese (Mn)-Oxygen (O) 
layers when the oxygens are moving out of the plane. 
}
\label{fig4}
\end{figure}

The discovery of a strong mixing down to the zone centre H=0 
shows that the long range properties of the system are affected. 
In explaining these findings, we thus propose to consider a 
coupling of the spin subsystem with atomic displacements within 
the unit cell \cite{19}. This in turn implies a hybridization 
mechanism with an optical phonon, as in the dynamical 
magnetoelectric coupling theory developed for orthorhombic 
$RMnO_3$ \cite{4,5,24}. In this scenario, the spin current 
$J_{ij} = S_i \times S_j$ (defined for neighbouring manganese 
spins $S_i$ and $S_j$ sitting at distance $r_{ij}$) plays a 
crucial role. Owing to the spiral magnetic ordering typical 
of these materials, $J_{ij}$ acts, via the inverse 
Dzyaloshinskii-Moryia coupling, as a force pushing the oxygen 
atom located between adjacent manganese off the Mn-Mn bond. 
This mechanism gives rise at $T_N$ to an electric polarization 
$P = r_{ij} \times J_{ij}$ lying within the spiral plane and 
perpendicular to $r_{ij}$. The multiferroic Goldstone mode 
is predicted to be a hybrid mode, rising at $\Delta_{SG}$, 
and made of a mixing between the optical phonon associated 
with the oxygen displacement, and the spin wave mode involving 
spin fluctuations out of the spiral plane. At first glance, 
this mechanism would not hold in the hexagonal case, since 
due to the triangular symmetry, the resulting magnetic force 
experienced by oxygen atoms is zero: indeed, each oxygen ion 
is located at the centre of a triangle formed by 3 neighbouring 
Mn ions. The system can however benefit from the 
Dzyaloshinskii-Moryia interaction by spontaneously moving the 
oxygen atoms along the c-axis. This distortion is expected to 
create an electric polarization parallel to $J_{ij}$ and is 
accompanied by a slight rotation of the spins towards the 
same direction. In that case, the coupled spin and atomic 
motions look like those of the ribs of an umbrella that would 
be put up or down (Figure \ref{fig4}). In close analogy with 
the orthorhombic case, hybridized spin-lattice Goldstone modes 
are expected at $\Delta_{SG}$, in agreement with the present 
results. 

To test the validity of this scenario, several predictions 
have to be further examined. First, the magnetic structure 
should be characterized by a tiny ferromagnetic moment 
superimposed on the AF structure, as sketched in the right 
panel of Figure \ref{fig4}. In that case, specific Bragg 
peaks should exhibit a specific contribution at $T_N$. 
Calculations of the structure factor show that this effect 
is best observed for Bragg peaks with a forbidden AF magnetic 
intensity and a very weak nuclear one. The (2-11) Bragg peak 
fulfils these conditions. As shown on Figure \ref{fig4}, its 
intensity increases below $T_N$, and this is a good indication 
for the validity of this scenario. Next, as the high temperature 
ferroelectric distortion is mainly due to atomic displacements 
parallel to the c-axis, the oxygen displacements proposed in 
this "umbrella scenario should result at $T_N$ in a slight 
change of the ferroelectric moment. Evidence for such an 
evolution has been recently reported by Lee et al \cite{19}, 
thanks to high resolution X rays and neutrons diffractions 
measurements. This result is another argument supporting our 
interpretation.

In conclusion, polarized inelastic neutron scattering 
experiments demonstrate the existence of hybrid spin and 
lattice low energy modes in $YMnO_3$, that can be considered 
as Goldstone modes of the multiferroic phase. The neutron 
polarization analysis directly shows their hybrid nature, 
revealing both spin and structural counterparts. The mechanism 
responsible for this hybridization could be the 
Dzyaloshinskii-interaction, in close analogy with the 
model recently proposed for orthorhombic multiferroic 
materials. 

\begin{acknowledgments} We would like to thank 
Yvan Sidis, Martine Hennion and Fernande Moussa 
for fruitful discussions.
\end{acknowledgments}

\end{document}